\documentclass[aps,onecolumn,superscriptaddress]{revtex4}
\usepackage{times}
\usepackage{color}
\usepackage{float}
\usepackage{epsfig}

\begin{document}

\title{From sparse to dense and from assortative to disassortative in online social networks }

\author{Menghui Li}\affiliation{Beijing Institute of Science and Technology Intelligence, Beijing, 100044, P. R. China} \affiliation{School of Systems Science,
Beijing Normal University, Beijing, 100875, P. R. China}
\affiliation{Temasek Laboratories, National University of Singapore, 117508, Singapore}
\affiliation{ Beijing-Hong Kong-Singapore Joint
Centre for Nonlinear \& Complex Systems (Singapore), National
University of Singapore, Kent Ridge, 119260, Singapore}

\author{Shuguang Guan}\email{guanshuguang@hotmail.com}
\affiliation{Department of Physics, East China Normal University, Shanghai, 200241, P. R. China}
\affiliation{State Key Laboratory of Theoretical Physics, Institute of Theoretical Physics, Chinese Academy of Sciences, Beijing, 100190, P. R. China}

\author{Chensheng Wu}\affiliation{Beijing Institute of Science and Technology Intelligence, Beijing, 100044, P. R. China}

\author{Xiaofeng Gong}
\affiliation{Temasek Laboratories, National University of Singapore, 117508, Singapore}
\affiliation{ Beijing-Hong Kong-Singapore Joint Centre for Nonlinear \& Complex Systems (Singapore), National University of Singapore, Kent Ridge, 119260, Singapore}

\author{Kun Li}
\affiliation{Temasek Laboratories, National University of Singapore, 117508, Singapore}
\affiliation{ Beijing-Hong Kong-Singapore Joint Centre for Nonlinear \& Complex Systems (Singapore), National University of Singapore, Kent Ridge, 119260, Singapore}

\author{Jinshan Wu}
\affiliation{School of Systems Science,
Beijing Normal University, Beijing, 100875, P. R. China}

\author{Zengru Di}
\affiliation{School of Systems Science,
Beijing Normal University, Beijing, 100875, P. R. China}

\author{Choy-Heng Lai}
\affiliation{ Beijing-Hong Kong-Singapore Joint
Centre for Nonlinear \& Complex Systems (Singapore), National
University of Singapore, Kent Ridge, 119260, Singapore}
\affiliation{Department of Physics,
National University of Singapore, 117542, Singapore}
\affiliation{Yale-NUS College, Singapore, 138614, Singapore}

\begin{abstract}
Inspired by the analysis of several empirical online social networks, we propose a simple reaction-diffusion-like
coevolving model, in which individuals are activated to create links based on their states, influenced by local dynamics and their own intention. It is shown that the model can reproduce the remarkable properties observed in empirical online social networks; in particular, the assortative coefficients are neutral or
negative, and the power law exponents $\gamma$ are smaller than $2$. Moreover, we demonstrate that, under appropriate conditions, the
model network naturally makes transition(s) from assortative to
disassortative, and from sparse to dense in their characteristics. The model is useful in
understanding the formation and evolution of online social networks.\\

\noindent{\bf SUBJECT AREAS:} COMPLEX NETWORKS, STATISTICAL PHYSICS, THEORETICAL PHYSICS, MODELLING AND THEORY
\end{abstract}
\maketitle

Massive websites --
Facebook, Twitter, MySpace, LinkedIn, Flickr, Orkut, Google+,
Weaklink, just to name a few -- are booming in the past few years,
where millions of users and their interactions naturally form the so called online social networks (OSNs)
\cite{Review,Review1,Review2}. For OSNs, one important characteristic is the strong interplay between the user behaviour and the network topology \cite{adaptive}. On the one hand, the
user behaviour is affected by the topology-dependent information
flowing in the networks
\cite{neighbor'effect1,neighbor'effect2,peerinfluence,socialinfluence}; on the other hand, the network topology continually evolves as a natural consequence of network dynamics
\cite{peerinfluence,dynamicaleffect,informationshapenetworks}.
Due to this feature, OSNs exhibit certain correlation patterns
during evolution, such as the highly skewed degree distributions
\cite{Weaklink,Google+,Orkut}, the generalized Gibrat's Law
\cite{Scalinglaws}, assortativity/disassortativity
\cite{Weaklink,Google+}, etc, which are of great importance for us
to understand the possible generic laws governing the organization
and evolution in networked systems \cite{Review-Oxford}.

Recently, two interesting phenomena in OSNs have attracted much attention.
The first one is related to the assortativity/disassortativity property of the network,
which is an important structural measure characterizing the degree correlation between pairwise nodes. Mathematically, the assortative coefficient can be defined as the Pearson correlation coefficient averaged for all pairs of adjacent nodes in the network. As
shown in Table \ref{table1}, it is reported that some OSNs (e.g., Twitter and Cyworld) show negative or neutral assortative coefficients \cite{Weaklink,Google+,Orkut,Twitter,Cyworld,pussokram,digg},
and some OSNs, such as Weaklink \cite{Weaklink}
and Google+ (G+) \cite{Google+}, even convert from being assortative to being disassortative during evolution.   These findings challenge our traditional knowledge \cite{mixing1,mixing} that biological and technical networks (e.g., financial networks \cite{financial})
are disassortative, while social networks (e.g., acquaintance networks \cite{Weighted}) are assortative.
Secondly, the scale-free property is of great importance for a network, which can be characterized by a power law exponent $\gamma$ as in $p(k)\sim k^{-\gamma}$, where $k$ and $p(k)$ are node degree and the distribution of degree, respectively. Under the thermodynamical limit, i.e., the network size $N \rightarrow \infty$, the mean degree of a scale-free network will diverge when $\gamma \leq 2$. Therefore,  $\gamma=2$ is an important boundary, and scale-free networks can be classified into dense ($\gamma \leq 2$) and sparse ($\gamma>2$) accordingly. Previously, many scale-free  networks are found to be sparse \cite{sparse}. However, as shown in Table \ref{table1},
some large OSNs, e.g., YouTube (YT), Digg, and
LiveJournal (LJ), turn out to be dense scale-free networks with $\gamma<2$ \cite{Orkut,digg,tianya}.

In Table \ref{table1}, the basic statistical properties for 14 popular OSNs are listed. It is found that these OSNs basically share common properties observed in real world networks, such as power-law distribution of degrees, large clustering coefficient, and small average shortest path. However, two features, i.e., negative or neutral assortative coefficients and $\gamma<2$, also turn out
to be typical. In order to obtain insights into the evolution patterns of real OSNs, it is desirable to set up a dynamical model which could reproduce the properties and dynamics observed in real OSNs. Previously, the power law distribution of degrees
\cite{Review,Review1,Review2,BAmodel,BBVmodel,popularityVSsimilarity}
and the disassortative correlation \cite{Weaklink,disassortative}
have been separately studied in theoretical models, and in most models the exponents $\gamma$ of degree distributions are larger than 2 (see Table III of Ref. \cite{Review} and the corresponding
references).
Recently, some theoretical works discussed the relevant properties of networks with specific functions to determine the degree distribution of the nodes \cite{generator,Robustness,SmallM}. However, attentions have not been paid to the dynamical origin of dense and/or disassortative OSNs, especially the transition from assortative to disassortative during the evolution of real networks.

\begin{table}
\caption{Properties of typical OSNs, including the number of
nodes $N$, the average degree $\langle k \rangle$, the average
shortest path $\langle d \rangle$,  the exponent of power law for out-degree (in-degree) $\gamma_{out}$ ($\gamma_{in}$), the average clustering coefficient $\langle c \rangle$, and the assortative coefficient $r$,  which is defined as the correlation between out-degree and in-degree as the links in OSNs are directional. The empirical data sets
analyzed in this paper are also listed here, i.e., \emph{Flickr},
\emph{FriendFeed (FF)} , \emph{aNobii}, and \emph{Epinions}. } \label{table1}
\begin{tabular}{@{\vrule height 10.5pt depth4pt  width0pt}lrcccccccc}
 Network&$N$&$\langle k \rangle$&$\langle d \rangle$&$\gamma_{out}(\gamma_{in})$&$\langle c \rangle$&$r$ \\
\hline
\emph{\textbf{Flickr}} & 2302925 &14.4& 5.7 & 1.75(1.74) &0.11& ~~0.02 \\

\emph{\textbf{FF}} & 204458 &20.6& 4.0 & 2.29(2.17) &0.19&$-0.10$ \\

\emph{\textbf{aNobii}} & 94238 &8.07& 5.3 & 2.71(2.70) &0.13&$-0.05$ \\

\emph{\textbf{Epinions}} & 114467 &5.63& 4.9 & 1.75(1.72) &0.08&$-0.06$ \\
\hline
Twitter \cite{Twitter} &470040&87.1 & - & 2.42(2.85) &0.11& $-0.26$\\

Cyworld \cite{Cyworld} &12048186&31.7 & 3.2 & - &0.17&$-0.13$\\

Nioki \cite{pussokram} & 50259&8.07& 4.1 & 2.2(2.4) &0.01&$-0.10$\\

Wealink \cite{Weaklink} & 223482&2.53& - &$2.91$ &-& $-0.07$\\

YT \cite{Orkut} & 1157827 &4.29& 5.1 & 1.63(1.99) & 0.14 & $-0.03$\\

Digg \cite{digg} &685719&9.8 & 5.6 & 1.6(1.5) &-&$-0.03$\\

G+ \cite{Google+} & 30000000&16& 6.9 &- &0.25& $-0.02$ \\

Tianya \cite{tianya} & 411554&-& - & 1.66 &0.07&~~0.03\\

Orkut \cite{Orkut} & 3072441 &106& 4.3 & 1.50(1.50) &0.17&~~0.07 \\

LJ \cite{Orkut} & 5284457 &17& 5.6 & 1.59(1.65) &0.33&~~0.18\\
\hline
\end{tabular}
\end{table}

Recently, we proposed a dynamical model based on empirical analysis of real OSNs such as Flickr and Epinions. It is shown that this simple reaction-diffusion-like model can reproduce statistical properties consistent with real data \cite{Li}. In this present paper we investigate, through modeling and simulations, the two remarkable observations that some OSNs are dense and/or disassortative.
Specifically, based on extensive empirical analysis of real network data, including Flickr, FriendFeed (FF), ANobii, and Epinions, we set up an evolution model, aiming at reproducing the above properties observed in OSNs. In the model, we characterize the user behaviour (local dynamics) in the OSNs by a state function. By considering a mechanism of local interplay generating new links, i.e., the formation of triadic closure, we are
able to describe the network evolution as a reaction-diffusion-like process, in which the network dynamics and topology evolve simultaneously and interdependently. As a natural consequence of the
coevolution, the resulting networks exhibit the
typical properties observed in real OSNs. Specifically, we show that the network is capable of making the transition from being sparse to
dense, and from being assortative to disassortative during the evolution.
We also offer some heuristic explanation for the above behaviour of the OSNs in our model.

Although the current model shares the same framework as in Ref. \cite{Li}, i.e, based on the reaction-diffusion-like local interaction pattern,  we emphasize that there are differences between them. Firstly, the modeling in Ref. \cite{Li} deals with typical dual-component networks consisting of users and items, while in this work, we only consider the social network, i.e, the user connections in the OSNs. Secondly, Ref. \cite{Li} mainly investigated
how the user connections, i.e., the social network, influence the formation of cross links (connections between users and items), and the dynamical correlations and patterns among different types of  degrees. In this paper, we focus on the dynamical origin of the transition from assortative to disassortative, and from sparse to dense in the OSNs characteristics. In addition, in the current model, we introduce the general Fermi function to simulate the diversity of user dynamics, which should be more reasonable than the random connection in Ref. \cite{Li}.


\begin{figure}
\begin{center}
\includegraphics{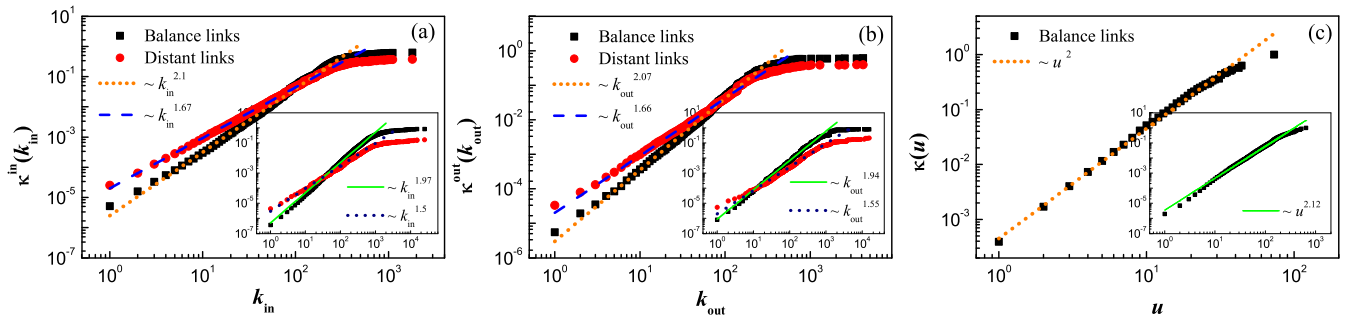}
\caption{The influence of the current topological
status on the formation of \emph{balance} links and \emph{distant}
links in the aNobii and FF (in the insets) networks. (a) The
cumulative functions of the relative probability $\kappa^{in}(k_{in})$
for PA versus the in-degree of the destination nodes;
(b) $\kappa^{out}(k_{out})$ for preferential creation versus the
out-degree of the source nodes; (c) The cumulative functions of the relative
probability $\kappa(u)$ for a pair of users to build a social link
given that they have already shared $u$ common neighbours for all
\emph{balance} links. The exponents are obtained by fitting the
curves of $\kappa (k)$ averaged over different initial snapshots.
The straight lines are guide to the eye throughout this
paper.}\label{fig1}
\end{center}
\end{figure}

\noindent\\ \textbf{Results}\\
{\bf Empirical analysis.} The mechanism of link formation is the central dynamical process during network evolution. In the
seminal work, Barab\'{a}si and Albert proposed a general rule governing the growth of networks, the preferential attachment (PA), which can explain the scale-free properties observed in many real world networks \cite{Review,Review1,Review2}.
Since then, much attention has been paid to the investigation of possible microscopic mechanisms underlying the PA phenomenon
\cite{Review,Review1,Review2}. So far, this important
question is still open and challenging. In this paper, we first
carry out empirical study on four typical OSNs, including Flickr
\cite{Flickr,Flickr1}, FF \cite{FriendFeed,FriendFeed1},
aNobii \cite{aNobii}, and Epinions \cite{Epinions} (see
\emph{Methods} for data description). Our particular interest is
on the patterns of link creation during network evolution.

To facilitate the analysis, we divide the new links into two mutually-exclusive types: the \emph{balance} links and the
\emph{distant} links based on the topological distance
\cite{Review-AMR}. If a new link is formed between a user and one of his second neighbours, i.e., the user who is two hops apart from him in the network, it is regarded as a \emph{balance} link \cite{Review-AMR}.
Otherwise, it belongs to the \emph{distant} links. Obviously, generating a \emph{balance} link always contributes a triangle in the network. By distinguishing between these two types of new links, we can investigate the dependence of new links on the topological distance.

\begin{table}[th]
\caption{Exponents $\alpha$ for empirical networks, characterizing the dependence of \emph{balance} links and \emph{distant} links (in the parentheses) on the degree and the number of common neighbours, i.e.,  $\kappa (x) \sim x^{\alpha+1}$. Here $\alpha_{A}$ for PA,
$\alpha_{C}$ for preferential creation, and $\alpha_{N}$ for common neighbours. For comparison, exponents $\alpha$ for \emph{balance} links in the model networks are also listed in the brackets.} \label{table2}
\begin{tabular}{@{\vrule height 10.5pt depth4pt  width0pt}lrccccc}

 Exponents&  Flickr &FF  & aNobii  & Epinions \\
\hline
$\alpha_{A}$ & 1.0 (0.48)[0.98]& 0.97 (0.5) [0.99]& 1.13 (0.70) [0.96] &1.22 (0.84)[ 0.97]\\

$\alpha_{C}$ & 1.0 (0.5)[1.19 ] & 0.9 (0.55)[0.96]& 1.11 (0.73)[0.94] &1.13 (0.56) [1.11]\\

$\alpha_{N}$ & 0.9 [1.14] & 1.12 [1.13] & 1.0 [0.95] & 0.95[1.13 ]\\
\hline
\end{tabular}
\end{table}

The main method we use to analyze the pattern of link growth is to measure the conditional probability that nodes acquire (create) new links with respect to their existing in-degree (out-degree)
\cite{PAmeasure,Local} (see \emph{Methods} for
details). The main empirical results for the four OSNs are summarized in Table \ref{table2} and illustrated in Fig. \ref{fig1}.
Interestingly, the relative probabilities of acquiring or creating new links satisfy a power law with respect to the existing degrees, indicating that the users with larger out-degree (in-degree) are more likely to
create (acquire) new links. Moreover, it is found that the exponents $\alpha$ for the \emph{balance} links are significantly larger than that
for the \emph{distant} links, as shown in Figs. \ref{fig1}(a) and \ref{fig1}(b). This suggests that the \emph{balance}  links depend on the local topological structure more than the
\emph{distant} links. We attribute this preferential formation of \emph{balance} links in the OSNs to the locality of information in such networks, i.e., usually users within a neighbourhood tend to influence each other.

To further examine the micro-dynamics in the process of link
formation, we measure the probability of forming \emph{balance} links with respect to the number of common neighbours between the source
node and the destination node. As shown in Table \ref{table2} and Fig.
\ref{fig1}(c), the probability is (approximately) linearly proportional to the
number of common neighbours. Thus the preferential
formation of \emph{balance} links can be understood as a two-step
random walk in the network. Here, by carefully examining the four
OSNs mentioned above, we obtain empirical evidence that the
preferential formation of triadic closure, i.e., the formation of
\emph{balance} links, can be one possible micro-dynamical process leading to the PA phenomenon in OSNs.

\begin{figure}
\begin{center}
\includegraphics{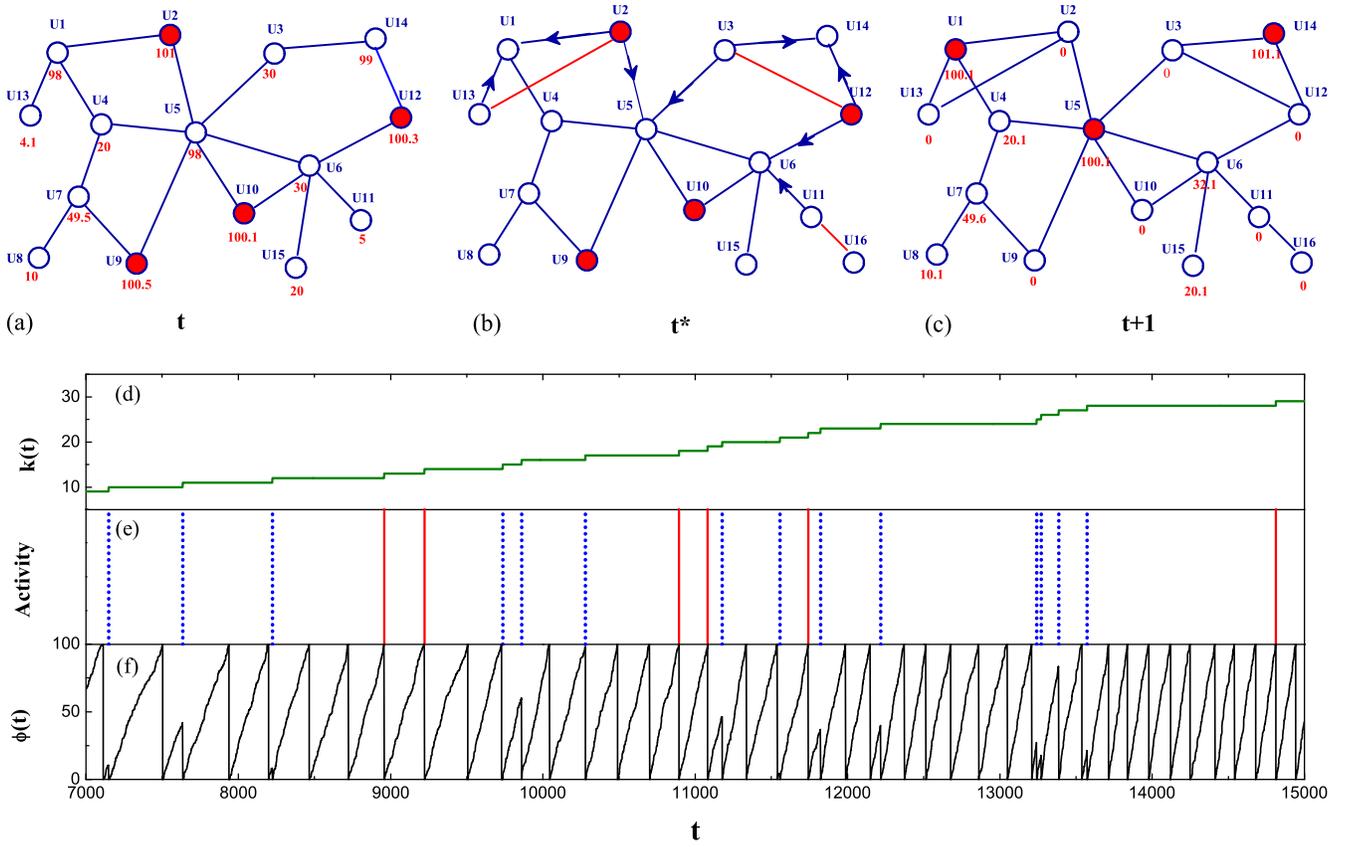}
\caption{Schematic illustration of the coevolution of both topology (a-c), and dynamical states (d-f) in the model. The numeric tags are the values of the state functions. (a) The network at time $t$, when some users (solid) are activated according to their states. (b) One step updating of the network topology: New user U16 joins and randomly connects to user U11; Some activated users connect to their second neighbours according to the acceptance probability $F(k)$, e.g., U2 to U13 and U12 to U3. The arrows represent the diffusion process. (c) At time $t+1$, the states of users are updated according to equation (\ref{state}). The states of activated users (U9 and U10) and those users building new links (U2, U13, U3, U14, U11, U16) at time $t$ are reset to $0$, but some nodes are activated again according to their states at time $t+1$. (d)-(f) Illustrating the evolution of the degree and the state function for a specific user during certain time period in the model. (d) Evolution of the degree $k(t)$. (e) The activities of the user. The solid lines indicate the moments when the user initially increases his social degree (e.g., U2 to U13 in (b)), and the dotted lines represent the moments when the user passively increases his social degree (e.g., U3 was connected by U12 in (b)), respectively. (f) Evolution of the state function $\phi (t)$. Parameters for the model:  $m=10, \mu=1, \phi_0=0.1, \Theta =100$. }\label{fig2}
\end{center}
\end{figure}

{\bf Modelling.} The above empirical analysis has shown that in the OSNs studied, user behaviour is essentially influenced by each
other within the neighbourhood, and such an interplay in turn regulates the global evolution of the network. This suggests that local dynamics plays a leading role in the formation of new links during evolution. Based on this finding, in the following we set up a coevolving model, which is only driven by local interactions at the microscopic level, i.e., preferential formation of triadic closures \cite{Vazquez2003,triadic,dilemmagame}  and influence within neighbourhood. For simplicity, we neglect the link directions in the modelling, i.e., we only consider an undirected network.

In order to describe the dynamics of the users, we introduce a state function $\phi(i,t)$ for each user in the network. Here $i$ and $t$ denote
the nodes and time, respectively. The values of the state functions
describe the willingness of the users to create links. For each user in the network, we assume that his state function satisfies the
following reaction-diffusion-like equation:
\begin{equation}\label{state}
\phi(i,t+1)-\phi(i,t)=\phi_0  + \mu \sum_j^{N}
a_{ij}[k_{j}(t+1)-k_{j}(t)],
\end{equation}
where  two parameters $\mu $ and $\phi_0$ are constants; $k_{j}(t)$ is the
degree of user $j$ at time $t$. The LHS of the equation is the
change of state function with time, which is driven by two "forces": reaction and diffusion. The first term on the RHS, i.e., $\phi_0$, is a source term denoting the reaction, which means that a user can change his state on his own. The second term describes the diffusion process, i.e., how the interplay in
the neighbourhood of the user $i$ changes his state function.  Basically, if the neighbours of user $i$ build new links, his state function will be increased as a result of this influence. We set a threshold $\Theta$ for the state function of each user. If the
state function exceeds the threshold, the user will be activated,
and has a probability $F(k_i)$ to actively create a new link. Once
a user has built a new link or his state function has exceeded the
threshold, his state $\phi(i,t)$ is reset to zero at the next time
step. Essentially, the model simulates the
user logins and activities in the OSNs in terms of the state functions.

\begin{figure}
\begin{center}
\includegraphics{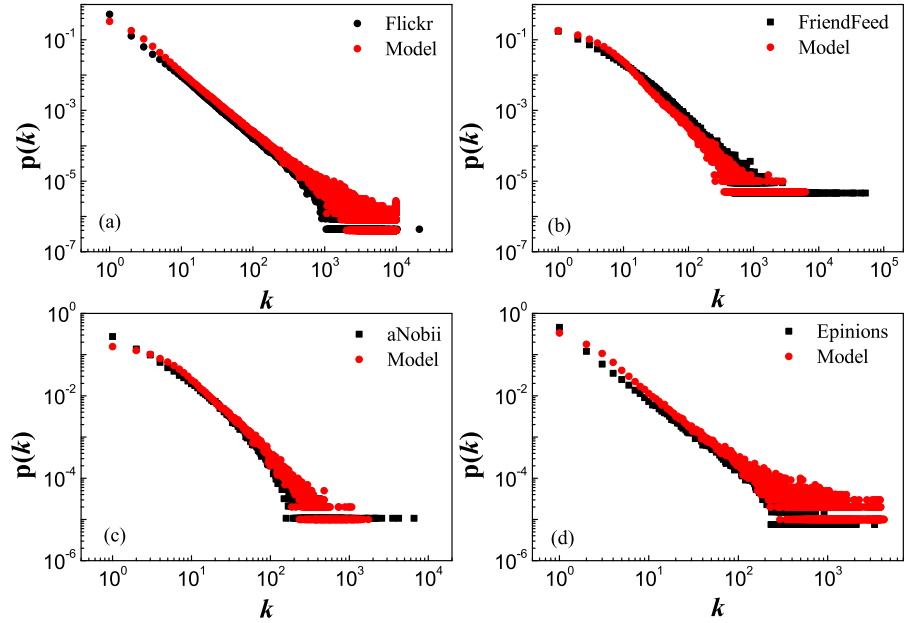}
\caption{Comparing the degree distributions of the empirical networks with that of the model network. Since the model network is undirected, we ignored the direction of links in the empirical networks for comparison.  (a) Flickr, where the parameters of the model are $m=25$, $N=500,000$, $\mu=1$, $\phi_0=0.01$, $\Theta=100$. (b) FriendFeed, where the parameters of the model are $m=8$, $N=200,000$, $\mu=1$, $\phi_0=0.02$, $\Theta=200$. (c) aNobii, where the parameters of the model are $m=7$, $N=100,000$, $\mu=1$, $\phi_0=0.02$, $\Theta=180$. (d) Epinions, where the parameters of the model are $m=25$, $N=100,000$, $\mu=1$, $\phi_0=0.01$, $\Theta=100$.}\label{fig3}
\end{center}
\end{figure}

As shown in Fig. \ref{fig1}(b), users with more
friends, i.e., with larger degrees, turn out to be more active in generating new links. To characterize the diversity of users' activities, we adopt a general Fermi function, which has been extensively used in evolutionary games models as the adaptive acceptance probability for each activated user \cite{games,dilemmagame}:
\begin{equation}\label{probability}
F(k_i)=\frac{1}{1+20e^{-0.001(k_i-\langle k\rangle)}}.
\end{equation}
Here $k_i$ is the degree of user $i$,   $\langle k\rangle = 2(m+1)$  is determined by the parameter $m$ in the model, representing the average degree of the whole network, and $0.001$ denotes the intensity of selection. $F(k_i)$ monotonically saturates to 1 with the increase of $k_i$, modulating the acceptance probability of nodes with different degrees. The parameter values (20 and 0.001) do not affect the qualitative behaviour of the model. In this paper, we choose the parameter values to allow the assortative coefficient vary in a relatively wide range.  We emphasize that the acceptance probability $F(k)$ may take other forms as long as it has similar behavior as the Fermi function.

Specifically, the algorithm to realize the model works as
follows: (1) At the very beginning, the initial network consists of a few users ($N_0$), forming a small connected random network. The state functions of users in the network evolve according to Eq.
(\ref{state}). (2) Adding users: at every time step, one new user is added to the network and randomly connects to an existing user. (3) Adding links: at each time step, $m$ users are randomly selected from the activated users  with the acceptance probability $F(k_i)$ (Eq. (\ref{probability})), and each connects to one of his second neighbours if they are not connected. If the number of
activated users is less than $m$, the remaining users are
randomly chosen from the network.   The above procedure is schematically illustrated in Fig. \ref{fig2}, where the states and the topology coevolve for one step driven by the local dynamics. As shown in Figs. \ref{fig2}(d)-\ref{fig2}(f), with the increase of $k$, the period of the state $\phi(i,t)$ for a user could become smaller, indicating that the users with larger degrees are more frequently activated.


\begin{figure}
\begin{center}
\includegraphics{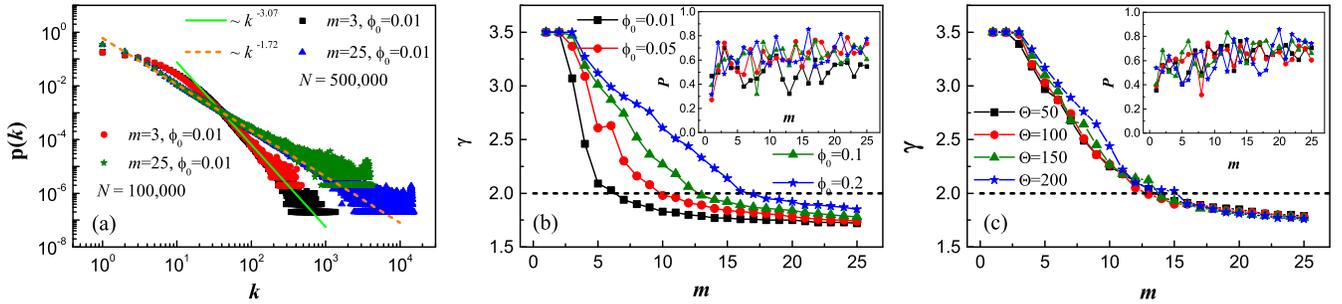}
\caption{Transition from sparse to
dense in the model network. (a) Degree distribution for $m=3$ and $m=25$ with different size $N$. (b)-(c) Power law exponents $\gamma$ with respect to parameter $m$ for different values of $\phi_0$ (b), and for different values of $\Theta$ (c). The insets are the p-values from the maximum likelihood method.
If not specified, the parameters in our simulations are $N=500,000$, $\mu=1$, $\phi_0=0.1$, $\Theta=100$ throughout the paper.  Results are averaged over 10 realizations. }\label{fig4}
\end{center}
\end{figure}

{\bf  Verifications.} In our model, although
we consider only simple local rules as the force driving
network evolution, numerical experiments have shown that the model can exhibit the main properties observed in empirical OSNs, such as the large clustering coefficient, small average shortest
path, and the power-law distributions of degrees, etc.    In order to verify our model, we first compare the degree distributions of the model network with that of the empirical networks in Fig. \ref{fig3}. It is found that the distributions are qualitatively consistent with each other under appropriate parameters. In empirical networks, the probability to build a new link depends on the existing degrees, as shown in Fig. \ref{fig1} and Table \ref{table2}. To compare the dynamics of our model with that of the empirical networks, we also applied the same analysis to the model under the same parameters of Fig. \ref{fig3} and summarized the results in Table \ref{table2} (in the brackets). It is seen that the characteristic exponents $\alpha$ are qualitatively
consistent with the empirical ones.

\begin{figure}
\begin{center}
\includegraphics{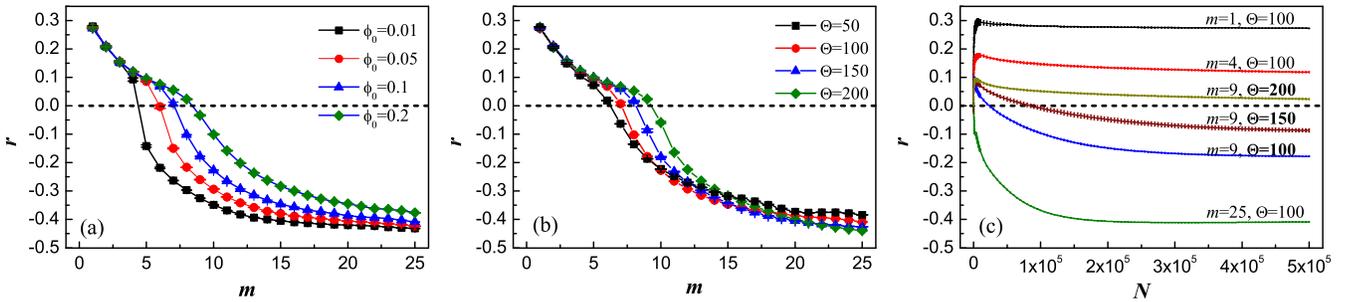}
\caption{Transition from assortativity to disassortativity in the model. (a)-(b) The assortative coefficient $r$ with respect to the parameters $m$ for different values of $\phi_0$ (a),  and for different values of $\Theta$ (b). The coefficients $r$ are calculated at the final stage in the model when $N=500,000$. (c) The temporal evolution of the assortative coefficient $r(N)$ for different $m$ at $\Theta=100$, and for different values of $\Theta$ at $m=9$. The error bar is the standard deviation. }\label{fig5}
\end{center}
\end{figure}

We now focus on the two major properties of the model network: the power-law exponent $\gamma$ and the assortative coefficient $r$. First, we investigate how the exponent $\gamma$ varies with respect to the model parameter $m$. In this work, the best power-law exponents $\gamma$ are calculated using the maximum likelihood method \cite{plfit}. As shown in Fig. \ref{fig4} (a), for small parameter $m$, the
distribution of degrees follows a stretched power law with the
exponents $\gamma$ larger than 2; while for large $m$, the exponent $\gamma$ turns out to be smaller than 2. As we know, many real world OSNs are characterized by $\gamma<2$. The present model can produce this important feature in flexible parameter regimes. In Fig. \ref{fig4} (a), we show the degree distributions for different network sizes. It is found that they are almost the same, indicating that the statistical properties of the model network are stable after long time evolution. We further find that, as parameter $m$ increases, the exponents $\gamma$ go down across $2$, as shown in Figs. \ref{fig4}(b) and \ref{fig4}(c), indicating that the generated network makes a transition from a sparse scale-free network to a dense network \cite{sparse}.   To justify the power law fitting, we compute the p-value for the power law model,
which measures how good the power law fitting is suitable for the data \cite{plfit}. As shown in the insets of Figs. \ref{fig4}(b) and \ref{fig4}(c), the p-values are generally larger than $0.25$  and the averages are 0.60 and 0.63, respectively, indicating the power law model is a plausible fit to the data.

We then investigate the assortative coefficient $r$ in the model \cite{mixing}. Since the links in the model are undirected, the assortative coefficient $r$ is defined as the correlation between degrees of pairwise nodes.
As shown in Figs. \ref{fig5}(a) and
\ref{fig5}(b), with the
increase of parameter $m$,  $r$ changes from positive
to negative, indicating that the model networks convert from being
assortative to being disassortative. There are two important points to emphasize. First, as shown in Fig. \ref{fig5}(a), the
change of the sign of $r$ occurs at larger $m$ as parameter $\phi_0$ increases. Second, as shown in Fig. \ref{fig5} (b), the value of $\Theta$ has significant influence on $r$.

In the above, we have shown that $r$ in the model could convert from positive to negative when parameter varies. As reported in Refs. \cite{Weaklink,Google+}, some OSNs convert from being assortative to being disassortative during evolution. How does $r$ in the model behave with the increase of time in our model?
First we note that the final network size $N$ is proportional to the total evolution time. As shown in Fig. \ref{fig5}(c), the
coefficients $r$ become almost stationary when the model
evolves for sufficiently long time.  In particular, in certain parameter
regimes, the generated networks evolve from the
initial assortativity to the subsequent disassortativity
with the increase of time. Therefore, the current model can characterize the distinct dynamical stages observed in the OSNs such as Weaklink and Google+ \cite{Weaklink,Google+}.

\begin{figure}
\begin{center}
\includegraphics{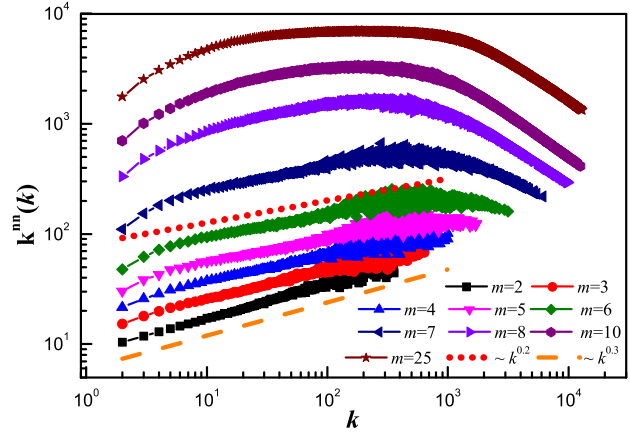}
\caption{Characterizing the average nearest
neighbours' degree $k^{nn}(k)$ for different values of the parameter $m$. The
corresponding assortative coefficients are 0.21, 0.15, 0.12, 0.09,0.07, 0.007, -0.10, -0.23, and -0.41 for increasing $m$, respectively.}\label{fig6}
\end{center}
\end{figure}

The assortative to disassortative change in our model can be heuristically understood based on Eq. (\ref{state}). Basically, it is the result of the competition between two factors in our model: the reaction factor denoted by parameter $\phi_0$, and the diffusion
factor denoted by parameter $\mu$. Parameter $m$ is important because it controls the diffusion and thus can change the ratio of these two factors. When $m$ is small, i.e.,
the number of new links formed at each time step is small, the
local influence is weak due to the small
average degree, i.e., $\langle k\rangle = 2(m+1)$.
In this case, the factor of reaction is relatively more important, and the user's own motive plays a dominant role in the evolution of the state function. Consequently,
the activation probability of a user is almost independent of the degree. Users thus have almost equal chance to be activated and connect to others, leading to the assortative
mixing pattern. This may correspond to the situations in some OSNs where users tend to establish links with people they know in real life, resulting in assortativity in the acquaintance network during the initial stage.
On the other hand, when $m$ is large, according to Eq. (\ref{state}), the local influence, i.e., the diffusion,  then plays a dominant role in the evolution of state function. In this case, users with
larger degrees have more chance to be activated and connect to others, leading to the disassortative mixing pattern.
In real situations, this may correspond to certain OSNs where the celebrities attract their fans to connect to them.

To further illustrate how parameter $m$ regulates the assortative mixing pattern in the model network, we calculate the average nearest neighbours' degree
$k^{nn}(k)$ in the generated networks \cite{Weighted}. As shown in Fig. \ref{fig6}, it is seen
that $k^{nn}(k)$ increases with respect to degree $k$ for small $m$, corresponding to positive assortativity in model networks. This is consistent with the situation in acquaintance
networks\cite{Weighted}. However, for larger $m$, $k^{nn}(k)$ increases first, and then decreases when the degree is large enough, corresponding to neutral and negative assortativity, as in some real OSNs \cite{Google+,digg,Orkut}. Similarly, the above analysis can also explain the results shown in Fig. \ref{fig5}(a), where a larger $m$ is required for the transition of $r$ when $\phi_0$ increases. Since $\phi_0$ represents the reaction factor, to overcome the outcome of increasing $\phi_0$ in the model, the diffusion factor needs to increase too. As a result of this competition, we observe that the transition occurs at a larger value of $m$.

In the evolution of real OSNs, generally the average
degree increases with time \cite{Google+,Douban}. This roughly corresponds to the increase of $m$ in the present model due to $\langle k\rangle=2(m+1)$.
As shown by our model in Fig. \ref{fig5}(c), this will cause the diffusion factor gradually to be dominant, and the network may convert from being assortative to being disassortative with the increase of time. Similarly,  the decrease of parameter
$\Theta$ is equivalent to the increase of parameter $m$, and the behaviour of the
model in Fig. \ref{fig5}(c) can also be explained from the viewpoint of competition between reaction and diffusion factors.

\begin{figure}
\begin{center}
\includegraphics{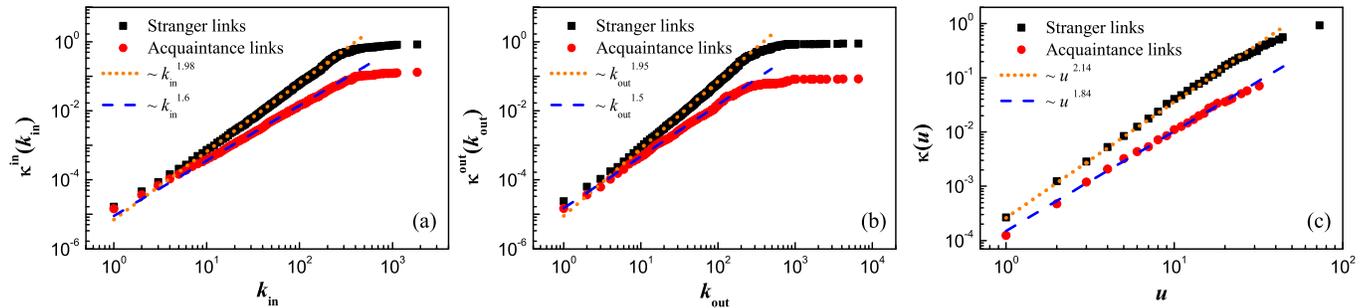}
\caption{Characterizing the difference between the acquaintance and stranger links in the aNobii network.  (a) The cumulative functions of the relative probability $\kappa^{in}(k_{in})$ versus in-degree of destination nodes; (b) $\kappa^{out}(k_{out})$ versus out-degree of source nodes. (c) The cumulative functions of the relative probability $\kappa(u)$ for a pair of users to build a social link given that they have already shared $u$ common neighbours.
}\label{fig7}
\end{center}
\end{figure}

To support our argument above, we apply empirical analysis to the aNobii network. Specifically, we regard it as a hybrid of a real world social network and a virtual online network. The former subnetwork consists of the \emph{acquaintance} links connecting users  knowing each other in real life, e.g., their family members and friends, such as ``Acquaintances" in Google+ and ``Friendship" in aNobii; and the latter comprises the \emph{stranger} links connecting their online virtual friends, such as ``Following" in Google+ and ``Neighbourhood" in aNobii. In terms of the reaction-diffusion process,
the generation of these two types of links is mainly due to
the reaction factor (i.e., user's personal desire) and the diffusion factor (i.e., the local influence) respectively.
Interestingly, we find that the subnetwork consisting of the
\emph{acquaintance} links is assortative with $r=0.06$,
like real world social networks.
On the contrary, the subnetwork consisting of the \emph{stranger} links is disassortative with $r=-0.09$. As shown in Fig. \ref{fig7}, the relative probabilities forming \emph{stranger} links are significantly larger than that forming \emph{acquaintance} links, implying that the diffusion factor is dominant in aNobii. As a result, the aNobii network as a whole turns out to be disassortative with $r=-0.05$.
The above results provide empirical evidence that the competition between diffusion and reaction might determine the mixing pattern of degrees in an OSN. Reasonably, during the evolution of the OSNs, if the diffusion factor dominates over the reaction factor, a transition from assortativity to disassortativity could be expected as in Weaklink \cite{Weaklink} and Google+ \cite{Google+}.


\noindent\\
\textbf{Discussion}\\
In this work, based on some empirical analysis of four typical OSNs, we set up a reaction-diffusion-like model, in which the evolution of the network is governed by both the users' personal motives and the influence within neighbourhood.
As a natural consequence of the coevolution of dynamics and topology, the model is able to qualitatively reproduce the major properties observed in real world OSNs. In particular, the generated networks can convert from being sparse to dense, and from being assortative to dissassortative with appropriate parameters. The model provides explanations of these two important features in real world OSNs in terms of the competition between reaction and diffusion factors in network evolution.

We believe that the current work is enlightening in modeling the evolution of the OSNs as well as of other real world networks. For example, other mechanisms of link formation, such as collective action and the structural hole mechanism, etc \cite{Review-AMR}, can be readily formulated and investigated. The idea of the model might be applicable to a wide range of social networks, and can be easily generalized to treat multi-layer networks, weighted networks, and social-attribute networks, etc. For example, recently, we have carried out a modeling for Flickr, with
a typical dual-component and dual-connection OSN, and obtained
satisfactory results \cite{Li}.

\noindent\\\textbf{Methods}\\
{\bf Data description and notations.}
\emph{\textbf{Flickr}} is one of the most famous websites sharing photos. The data set for our study is collected by daily crawling the Flickr network over 2.3 million users from Nov 2, 2006 to Dec 3, 2006, and again daily from Feb 3, 2007 to May 18, 2007. In total, there are 104 days in the time window of data collection \cite{Flickr,Flickr1} (http://socialnetworks.mpi-sws.org/). There are more than 2.3 million users and 33 million directed links among them.
\emph{\textbf{FriendFeed (FF)}} is a content aggregation site where users discover and discuss the interesting contents found
on the web by their friends. The data set is collected by
crawling the FriendFeed network once within every five days between
Feb 26 and  May 6, 2009 \cite{FriendFeed,FriendFeed1}. There are 14 snapshots or 70 days in the time window. More than 200 thousand users were found and about 4 million directed links among them were identified.
\emph{\textbf{ANobii}} is a website where readers can rate, review and discuss books with others.
The data set is collected by crawling the
neighbourhood (stranger in real life) and friendship (acquaintance in real life) networks of aNobii. Six snapshots of the network, 15 days apart, are collected starting from Sep 11, 2009 \cite{aNobii}. Users connect to each other through two mutually-exclusive types of ties: friendship and neighbourhood links. At last, the aNobii network includes 86,800 users, 429,482 \emph{stranger} links and 268,655 \emph{acquaintance} links.
\emph{\textbf{Epinions}} is a consumer review website where users can write reviews of products and also "trust" or
"distrust" each other. The data set contains the trusted relationships among users before Aug 12, 2003 \cite{Epinions}
(http://www.trustlet.org/wiki/Extended\_Epinions\_dataset),
including 114,467 users and 717,667 trusted relations. Mathematically, we can use the adjacency matrix $A_{N\times N}$ to characterize the topology of the online social networks, where $a_{ij}=1$ if user $i$ declares user $j$ as friend, otherwise $0$. Since the links among users are directional in these four networks, we accordingly define two types of degrees: the out-degree $k_{out}(i)=\sum_j a_{ij}$,
i.e., the number of friends claimed by user $i$, and the in-degree $k_{in}(j)=\sum_ia_{ij}$, i.e., the number of users who claim user $j$ as friend. The statistical properties of these four data sets are listed in Table \ref{table1}.

{\bf Measuring preferential attachment.} In Refs.
\cite{PAmeasure,Local}, a numerical method is used to measure the
preferential attachment (PA) growth of a network. Given that we know the
temporal order in which the nodes join the network, the essential
idea of the method is to monitor to which existing node the new nodes
connect, as a function of the degree of the old node. We take an
example to briefly explain the method as follows: (1) At time $t_0$,
we mark the nodes with $k_{out}$ out-degree as ``$t_0$ nodes",
denoting their number as $C(k_{out})$. (2) After the evolution of a period $\Delta t$, the out-degrees of the ``$t_0$ nodes" have increased due to the evolution of the network (of course, the in-degrees also change). We count the out-degree created by the ``$t_0$ nodes" as $A(k_{out})$. Since we divide the newly generated links into two types, i.e., the \emph{balance} and the \emph{distant}, we have $A(k_{out})=A_B(k_{out})+A_D(k_{out})$, where the subscripts $B$ and $D$ denote the two types, respectively.  (3) The histogram providing the number of out-degree acquired by the ``$t_0$ nodes"
with exact $k_{out}$ out-degree, after normalization, defines a
function:
\begin{equation}
\Pi_i^{out}(k_{out})
=\frac{A_i(k_{out})}{C(k_{out})}/\sum_{k_{out}'}\frac{A(k_{out}')}{C(k_{out}')},
\label{limh}
\end{equation}
where, $i$ can be either $B$ or $D$. It has been proven that if PA mechanism exists, the conditional probability with which the
out-degree grows with respect to the existing out-degree follows a power law, namely $ \Pi_i^{out}(k_{out})\propto k_{out}^{\alpha}$.
Numerically, it is convenient to examine the cumulative function of $\Pi_i^{out}(k_{out})$,  which will also follow a power law, i.e.,
\begin{equation}
\kappa_i^{out}(k_{out})=\int^{k_{out}}_{0}
\Pi_i^{out}(k'_{out})dk'_{out}
 \propto k_{out}^{\alpha + 1}.
\label{cumulative}
\end{equation}
Similarly, the above numerical method can be applied to  the calculation of the probability for acquiring new links with respect to in-degrees $k_{in}$, and the treatment of the \emph{acquaintance} and \emph{stranger} links is straightforward.

\noindent\\ \textbf{Acknowledgments} \\
We are particularly grateful to A. Mislove for sharing the Flickr database, P. Massa for sharing the Epinions database, A. Barrat for sharing the aNobii database, and T. Gupta for sharing the FF database. This work is sponsored by Technology Foundation for Selected Overseas Chinese Scholar. SGG is sponsored by the following funding agencies: Science and Technology Commission of Shanghai Municipality under grant No. 10PJ1403300; Innovation Program of Shanghai Municipal Education Commission under grant No. 12ZZ043; the Open Project Program of State Key Laboratory of Theoretical Physics, Institute of Theoretical Physics, Chinese Academy of Sciences, China (No. Y4KF151CJ1); and the NSFC under grant Nos. 11075056 and 11135001. This work is also supported by NSFC under Grant Nos. 61174150 and 61374175.

\noindent \\ \textbf{Author contributions} \\
M.H.L., S.G.G., C.S.W, X.F.G., K.L., J.S.W., Z.R.D. and C.H.L. designed research and analyzed the data;  M.H.L., S.G.G., and C.S.W. performed research and wrote the paper. All authors reviewed and approved the manuscript.

\noindent \\ \textbf{Additional information}\\
 Competing financial interests: The authors declare no competing financial interests.

\noindent Correspondence and requests for materials should be addressed to S.G.G. (guanshuguang@hotmail.com)



\begin{thebibliography}{99}

\bibitem{Review} Albert, R., \& Barab\'{a}si,  A. L.
Statistical mechanics of complex networks. {\it  Rev. Mod. Phys.}
\textbf{74} 47-97 (2002).

\bibitem{Review1} Newman, M. E. J.
The structure and function of complex networks. {\it SIAM Rev.}
\textbf{45} 167-256 (2003).

\bibitem{Review2} Boccaletti, S., Latora, V., Moreno, Y., Chavez, M.,
\& Hwang, D.-U.  Complex networks: Structure and dynamics. {\it
Phys. Rep.} \textbf{424} 175-308 (2006).


\bibitem{adaptive} Gross, T. \& Blasius, B.  Adaptive coevolutionary networks: a review.  {\it J. R. Soc. Interface} \textbf{5}  259-271 (2008).

\bibitem{neighbor'effect1} Centola, D.  The spread of behavior
in an online social network experiment. {\it Science} \textbf{329} 1194-1197 (2010).


\bibitem{neighbor'effect2} Onnela, J.-P. \&  Reed-Tsochas,  F.  Spontaneous
emergence of social influence in online systems. {\it Proc Natl
Acad Sci USA} \textbf{107} 18375-18380 (2010).

\bibitem{socialinfluence}  Bond, R. M. et al. A 61-million-person experiment in social influence and political mobilization. {\it Nature} \textbf{489} 295-298 (2012).

\bibitem{peerinfluence} Lewisa, K., Gonzalez, M. \& Kaufman, J. Social selection and peer influence in an online social network. {\it Proc Natl Acad Sci USA} \textbf{109}  68-72 (2011).


\bibitem{dynamicaleffect} Kossinets, G. \& Watts, D. J. Empirical analysis of an evolving
social network. {\it Science} \textbf{311}  88-90 (2006).

\bibitem{informationshapenetworks}Rocha, L. E. C., Liljeros, F. \& Holm, P. Information dynamics shape the sexual networks of Internet-mediated prostitution. {\it Proc
Natl Acad Sci USA} \textbf{107} 5706-5711 (2010).


\bibitem{Weaklink}Hu, H.-B. \& Wang, X.-F.  Disassortative mixing in online social networks. {\it Europhys. Lett.} \textbf{86} 18003 (2009).

\bibitem{Google+} Gong, N. Z. et al. Evolution of Social-Attribute Networks:
Measurements, Modeling, and Implications using Google+. {\it Proceedings of the 2012 ACM conference on Internet measurement conference (IMC'12)}, Boston, USA
pp 131-144; DOI:10.1145/2398776.2398792 (2012).


\bibitem{Orkut} Mislove, A., Marcon, M., Gummadi, K. P., Druschel, P.,
\& Bhattacharjee, B.  Measurement and analysis of online social
networks. {\it Proceedings of the 7th ACM SIGCOMM conference on Internet measurement (IMC'07)}, San Diego, CA, USA, pp 29-42; DOI:10.1145/1298306.1298311 (2007).

\bibitem{Scalinglaws} Rybski, D., Buldyrev, S. V., Havlin, S., Liljeros, F. \&
Makse, H. A.  Scaling laws of human interaction activity. {\it
Proc Natl Acad Sci USA} \textbf{106} 12640-12645 (2009).

\bibitem{Review-Oxford} Newman, M. E. J., Barab\'{a}si, A.-L. \& Watts, D. J.  \emph{The Structure and Dynamics of Networks} (Princeton University Press) (2006).

\bibitem{Twitter} Zhou, Z., Bandari, R., Kong, J., Qian, H. \& Roychowdhury, V. Information Resonance on Twitter: Watching Iran. {\it Proceedings of the First Workshop on Social Media Analytics (SOMA'10)}, Washington, USA, pp 123-131; DOI:10.1145/1964858.1964875 (2010).

\bibitem{Cyworld}  Ahn, Y., Han, S., Kwak,  H., Moon, S. \&  Jeong, H.  Analysis of Topological Characteristics of Huge Online
Social Networking Services. {\it Proceedings of the 16th international conference on World Wide Web (WWW'07)},   Banff, AB, Canada, pp 835-844; DOI:10.1145/1242572.1242685 (2007).

\bibitem{pussokram} Holme, P., Edling, C. R. \& Liljeros, F. Structure and time evolution of an Internet dating community. {\it Social Networks} \textbf{26} 155-174 (2004).

\bibitem{digg} Tang, S., Blenn, N., Doerr, C. \& Van Mieghem, P.  Digging in the Digg Social News Website. {\it IEEE Trans on Multimedia} \textbf{13} 1163-1175 (2011).

\bibitem{mixing1}  Newman, M. E. J. Assortative mixing in networks. {\it Phys. Rev. Lett.} \textbf{89} 208701 (2002).

\bibitem{mixing}  Newman, M. E. J.  Mixing patterns in networks. {\it Phys. Rev. E}
\textbf{67}  026126 (2003).

\bibitem{financial} Cai, S., Zhou, Y., Zhou, T. \& Zhou, P. Hierarchical organization and disassortative mixing of correlation-based weighted financial networks. {\it Int. J. Mod. Phys. C} \textbf{21} 433-441 (2010).

\bibitem{Weighted} Barrat, A., Barth\'{e}lemy, M., Pastor-Satorras, R. \& Vespignani,
A.  The architecture of complex weighted networks. {\it Proc
Natl Acad Sci USA}  \textbf{101} 3747-3552 (2004).

\bibitem{sparse} Genio, C. I. D., Gross, T.,
\& Bassler, K. E.  All scale-free networks are sparse. {\it Phys.
Rev. Lett.} \textbf{107} 178701 (2011).

\bibitem{tianya} Xiong, F., et al. A dissipative network model with neighboring activation. {\it Eur. Phys. J. B} \textbf{84} 115-120 (2011).




\bibitem{BAmodel} Barab\'{a}si, A. L. \& Albert, R.
Emergence of scaling in random networks. {\it Science} \textbf{286} 509-512 (1999).

\bibitem{BBVmodel} Barrat, A., Barth\'{e}lemy, M. \& Vespignani, A. Weighted Evolving Networks: Coupling Topology and Weight Dynamics. {\it Phys. Rev.
Lett.} \textbf{92} 228701 (2004).

\bibitem{popularityVSsimilarity} Papadopoulos, F.,
Kitsak, M., Serrano, M., Bogu\~{n}\'{a}, M. \& Krioukov, D. Popularity
versus similarity in growing networks. {\it Nature} \textbf{489}  537-540 (2012).

\bibitem{disassortative} Leung, C. C. \& Chau, H. F.  Weighted assortative and disassortative networks model. {\it Physica A} \textbf{378} 591-602 (2007).



\bibitem{generator} Palla, G., Lov\'{a}sz, L., \&  Vicsek T. Multifractal network generator. {\it Proc Natl Acad Sci USA} \textbf{107} 7640-7645 (2010).


\bibitem{Robustness} D'Agostino G., Scala A., Zlatic V. \& Caldarelli, G. Robustness and Assortativity for Diffusion-like Processes in Scale-free Networks. {\it EPL} \textbf{97} 68006 (2012).

\bibitem{SmallM} Small, M., et al. Scale-free networks which are highly assortative but not small world. {\it Phys. Rev. E} \textbf{ 77}  066112 (2008).

\bibitem{Li} Li, M., et al.  A coevolving model based on preferential triadic closure for social media networks. {\it Sci. Rep.}  \textbf{3}  2512 (2013).


\bibitem{Flickr} Mislove, A.,  Koppula, H. S., Gummadi, K., Druschel, P. \& Bhattacharjee, B. Growth of the Flickr Social Network. {\it Proceedings of the first workshop on Online social networks (WOSN'08)}, SEATTLE, WA, USA, pp 25-30; DOI:10.1145/1397735.1397742 (2008).

\bibitem{Flickr1} Cha, M., Mislove, A. \& Gummadi, K. P.
A Measurement-driven analysis of information propagation in the
Flickr social network. {\it Proceedings of the 18th international conference on World wide web (WWW'09)}, Madrid, Spain, pp 721-730; DOI:10.1145/1526709.1526806 (2009).

\bibitem{FriendFeed} Garg, S., Gupta, T., Carlsson, N. \& Mahanti, A.  Evolution of an Online Social Aggregation Network:
An Empirical Study. {\it Proceedings of the 9th ACM SIGCOMM
conference on Internet measurement conference(IMC'09)}, Chicago, USA, pp 315-312; DOI:10.1145/1644893.1644931(2009).

\bibitem{FriendFeed1} Gupta, T., Garg, S., Mahanti, A., Carlsson, N. \& Arlitt, M.  Characterization of
FriendFeed - A Web-based Social Aggregation Service. {\it
Proceedings of the Third International ICWSM Conference (2009)}, San Jose, California, pp 218-221. Menlo Park, California: AAAI Press (2009).

\bibitem{aNobii} Aiello, L. M., Barrat, A,, Cattuto,  C., Ruffo, G. \& Schifanella, R. Link creation and profile alignment in the aNobii
social network. {\it Proceedings of the Second IEEE International Conference on Social Computing (SocialCom 2010)}, Minneapolis, MN, USA, pp 249-256; DOI:10.1109/SocialCom.2010.42 (2010).

\bibitem{Epinions} Massa, P. \& Avesani, P.  Trust-aware bootstrapping of
recommender systems. { \it Proceedings of ECAI 2006 Workshop on
Recommender Systems}, Riva del Garda, Italy, pp 29-33 (2006).


\bibitem{Review-AMR} Contractor, N. S., Wasserman,  S. \& Faust, K.  Testing Multitheoretical,
Multilevel Hypotheses About Organizational Networks: An Analytic
Framework and Empirical Example. {\it Acad. Manage Rev.}  \textbf{31} 681-703 (2006).

\bibitem{PAmeasure} Jeong, H., N\'{e}da, Z., Barab\'{a}si, A. L. (2003)
Measuring preferential attachment in evolving networks. {\it
Europhys. Lett.}  \textbf{61} 567-572 (2003).

\bibitem{Local} Li, M., Gao, L., Fan, Y., Wu, J. \& Di, Z. Emergence
of global preferential attachment from local interaction. {\it New J.
of Phys.}  \textbf{12} 043029 (2010).

\bibitem{Vazquez2003} V{\'a}zquez, A.  Growing network with local rules: Preferential attachment, clustering hierarchy, and degree correlations.
{\it Phys. Rev. E} \textbf{ 67}  056014 (2003).

\bibitem{triadic} Klimek, P. \& Thurner, S.  Triadic closure dynamics drives scaling laws in social multiplex networks. {\it New. J. of Phys.}  \textbf{15}  063008 (2013).

\bibitem{dilemmagame} Szolnoki1, A., Perc, M. \& Danku,  Z. Making new connections towards cooperation in the prisoner's dilemma game. {\it EPL} \textbf{84} 50007 (2008).

\bibitem{games} Perc, M. \& Szolnoki, A. Coevolutionary games-A mini review. { \it BioSystems} \textbf{99} 109-125 (2010).

\bibitem{plfit}Clauset, A., Shalizi, C. R. \& Newman, M. E. J.  Power-Law Distributions in Empirical Data. {\it SIAM Review}  \textbf{51}  661-703 (2009).

\bibitem{Douban}Zhao, J., Luit, J. C. S., Towsley, D., Guan, X. \& Zhou, Y.
Empirical Analysis of the Evolution of Follower Network: A Case Study on Douban. {\it 2011 IEEE Conference on Computer Communications Workshops (INFOCOM WKSHPS),} Shanghai, China, pp 924-929; DOI:10.1109/INFCOMW.2011.5928945 (2011).



\end{thebibliography}
\end{document}